# Adaptive Gain and Order Scheduling of Optimal Fractional Order PI$^\lambda$D$^\mu$ Controllers with Radial Basis Function Neural-Network

Saptarshi Das, Sayan Saha, Ayan Mukherjee, Indranil Pan, Amitava Gupta

*Abstract*—Gain and order scheduling of fractional order (FO) PI$^\lambda$D$^\mu$ controllers are studied in this paper considering four different classes of higher order processes. The mapping between the optimum PID/FOPID controller parameters and the reduced order process models are done using Radial Basis Function (RBF) type Artificial Neural Network (ANN). Simulation studies have been done to show the effectiveness of the RBFNN for online scheduling of such controllers with random change in set-point and process parameters.

## I. Introduction

IN process control applications, most process models are approximated by reduction to various standard templates like First Order Plus Time Delay (FOPTD), Second Order Plus Time Delay (SOPTD) etc. Hence, controller tuning rules for these standard templates is essential for the effective performance of the plant and is very useful from the operator's view-point. A time-varying plant can be approximated at various instants of time by these standard templates and the corresponding gains of the controller may be updated online to efficiently control the plant. This method is commonly a form of adaptive control known as gain scheduling of PID controllers and is widely used in the process industries as an effective means to compensate for variations in process parameters [1]-[5]. Another technique to deal with such kind of time varying processes is the design of robust controllers which work well for a wide range of operating conditions [6]. However in such cases the performance of the nominal system is compromised although the controller implementation is relatively easy.

The gain scheduling itself may be done by various methods like a lookup table, an analytic expression which relates the auxiliary variables to the process parameters, neural network based function approximators, manual tuning by the operator etc [7]. In [8] neural networks have been used for online updating of PID controller parameters for non-linear process control applications. Conventionally, a neuro-emulator is trained off-line to mimic the non-linear process and is linearized at each sampling point as in [8]-[10]. The output of the neuro-emulator is fed to the neuro-tuner to get the updated PID controller parameters thus producing an intelligent PID controller [11]-[12]. RBF neural network has certain advantages over other ANN types in the design of adaptive PID control system [13]-[15]. Other applications of ANN based PID control like multilayer ANN containing PID algorithm [16] and ANN based robust PID control to ensure good performance for process delay variation [17] etc.

Performance of fractional order PID controllers has also been enhanced using ANN as in [18]. However, after the initial tuning process, only the gains are adaptively changed while the fractional orders are held constant. Radial Basis Function Neural Networks have been used in [19] to mimic complicated frequency domain tuning strategy so as to derive relationships between the plant parameters and the controller parameters and simplify the tuning process itself. Fractional order controllers have been used for gain and order scheduling to take care of performance degradation of network based control applications for handling variable delays in [20] and random packet-loss and delays with optimization based tuning in [21].

In this paper, a test-bench of higher order processes [22] are tuned with PID/FOPID controllers using Genetic Algorithm (GA) as in [23]. The corresponding gains of the PID/FOPID controller for different process parameters are obtained using an ANN based approach and are scheduled considering the processes to be time varying in nature. The time varying parameters of the linear models may be estimated online by recursive identification algorithms as done in traditional adaptive PID control designs [1]. Other process identification approaches may involve the use of an offline trained neural network which acts as an emulator as in [7]. The effectiveness of this methodology is enunciated with the help of extensive simulation studies. The methodology, presented in this paper is also capable of handling non-linear processes which can be linearized into standard reduced order templates around each operating condition and thus has wide industrial applicability.

Since the parameters of the controller are time varying, the analytical closed loop stability is very difficult to establish for all possible cases of process switching [24]. Asymptotic stability may be guaranteed for each operating conditions but very high switching transients might exist when the parameters of the controller are updated. To counter this, a hierarchical supervisor may be used to monitor the closed loop performance of the process and at the onset of instability corrective actions may be taken. Detailed discussions regarding the stability and performance of gain scheduling of PID controllers for time-varying processes have elucidated in [25]-[26].

Manuscript received April 15, 2011. This work has been supported by the Department of Science & Technology (DST), Govt. of India under the PURSE programme.

S. Das is with School of Nuclear Studies & Applications, Jadavpur University, Salt-Lake Campus, LB-8, Sector 3, Kolkata-700098, India. (E-mail: saptarshi@pe.jusl.ac.in).

S. Saha is with Dept. of Instrumentation and Electronics Engineering, Salt-Lake Campus, LB-8, Sector 3, Kolkata-700098, India.

A. Mukherjee, I. Pan and A. Gupta are with Dept. of Power Engineering, Jadavpur University, LB-8, Sector 3, Kolkata-700098, India.

The rest of the paper is organized as follows. Section II introduces with the family of higher order test-bench processes that are considered to be switched from one to the other. Section III reports GA based optimal PID and $PI^\lambda D^\mu$ controller tuning and the training of the RBFNN to map reduced SOPTD parameters to the optimum controller parameters. Superiority of gain and order scheduling of $PI^\lambda D^\mu$ controllers are shown in section IV over classical gain scheduled PID controllers. The paper ends with the conclusion as section V, followed by the references.

## II. THEORETICAL FORMULATION

### A. Sub-optimum Model Reduction for Higher Order Processes

Higher order process models are converted to SOPTD template by minimizing the discrepancy between the frequency responses of the higher order model $P(s)$ and reduced parameter process model $\widetilde{P}(s)$ in the Nyquist plane while minimizing the following objective function:

$$J_{nyquist} = w_1 \cdot \left\| \text{Re}\left[P(j\omega)\right] - \text{Re}\left[\widetilde{P}(j\omega)\right] \right\| + w_2 \cdot \left\| \text{Im}\left[P(j\omega)\right] - \text{Im}\left[\widetilde{P}(j\omega)\right] \right\| \quad (1)$$

This is an improved version of the $H_2$ norm based methodology proposed by Xue & Chen [27] since it minimizes the discrepancies in both the gain and the phase of the system and produces better accuracy for SOPTD templates given by:

$$P_{FOPTD}(s) = \frac{Ke^{-Ls}}{(\tau_1 s + 1)(\tau_2 s + 1)} \quad (2)$$

Here, the system parameters $\{K, \tau_1, \tau_2, L\}$ denotes the dc-gain, two time-constants and time-delay respectively.

### B. Test-Bench Processes

For testing our proposed algorithms a standard test-bench of higher order processes which are normally encountered in process control applications are considered [22].

$$P_1(s) = \frac{1}{(1+s)^n}, n \in \{3, 4, 5, 6, 7, 8, 10, 20\} \quad (3)$$

$$P_2(s) = \frac{1}{(1+s)(1+\alpha s)(1+\alpha^2 s)(1+\alpha^3 s)}, \quad (4)$$

$$\alpha \in \{0.1, 0.2, 0.3, 0.4, 0.5, 0.6, 0.7, 0.8, 0.9\}$$

$$P_3(s) = \frac{1}{(1+s)(1+sT)^2}, \quad (5)$$

$$T \in \{0.005, 0.01, 0.02, 0.05, 0.1, 0.2, 0.5, 2, 5, 10\}$$

$$P_4(s) = \frac{(1-\alpha s)}{(1+s)^3}, \quad (6)$$

$$\alpha = \{0.1, 0.2, 0.3, 0.4, 0.5, 0.6, 0.7, 0.8, 0.9, 1.0, 1.1\}$$

The four classes of process models are considered in this paper, which represents the dynamics of an arbitrary time-varying process and attempted to be efficiently handled by scheduling of optimally tuned PID controller parameters.

## III. TIME DOMAIN OPTIMAL CONTROLLER TUNING

### A. Controller Structures and Performance Index for Optimal Tuning

The $PI^\lambda D^\mu$ controller considered here has a parallel structure (7) like the conventional PID controller.

$$C_{FOPID}(s) = K_p + \frac{K_i}{s^\lambda} + K_d s^\mu \quad (7)$$

Clearly, the $PI^\lambda D^\mu$ controller (7) is a generalization of the classical PID controller with two extra tuning knob i.e. the differ-integral orders $\{\lambda, \mu\}$. By putting $\{\lambda, \mu\} = 1$, we can get the classical PID controller.

The PID and $PI^\lambda D^\mu$ controllers are now tuned with a constrained GA while minimizing the control objective ($J$) defined in (8). The goal of the constrained optimization is to minimize a weighted sum of an error index and the control signal, given by:

$$J = \int_0^\infty \left[w_1 \cdot t \cdot |e(t)| + w_2 \cdot u^2(t)\right] \quad (8)$$

Here, the first term corresponds to the Integral of Time multiplied Absolute Error (ITAE) which minimizes the overshoot and settling time, whereas the second term denotes the Integral of Squared Controller Output (ISCO). The two weights $\{w_1, w_2\}$ balances the impact of control loop error (oscillation and/or sluggishness) and control signal (larger actuator size and chance of integral wind-up) and both have been chosen to be unity in the present simulation study indicating same penalty for large magnitude ITAE and ISCO.

### B. Genetic Algorithm for Optimal Controller Tuning

Genetic algorithm (GA) is a stochastic optimization process which can be used to minimize a chosen objective function. A solution vector is initially randomly chosen from the search space and undergoes reproduction, crossover and mutation, in each iteration to give rise to a better population of solution vectors in the next iteration. Reproduction implies that solution vectors with higher fitness values can produce more copies of themselves in the next generation. Crossover refers to information exchange based on probabilistic decisions between solution vectors. In mutation a small randomly selected part of a solution vector is occasionally altered, with a very small probability. This way the solution is refined iteratively until the objective function is minimized below a certain tolerance level or the maximum number of iterations are exceeded. The number of population members in GA is chosen to be 20. The crossover and mutation fraction are chosen to be 0.8 and 0.2 respectively for the present simulation study.

## C. Radial Basis Function Neural Network Based Controller Parameter Adaptation

Artificial neural networks have been widely used in the field of control systems for the purpose of system identification, nonlinear modeling, gain-adaptation etc. [28]. Its wide applicability stems from the fact that the weights of the individual neurons can be trained with several algorithms (like back-propagation, GA etc.) so that the entire network can ultimately approximate almost any given non-linear function. There are no specific guidelines for choosing the number of hidden layers, bias weights, choice of interconnections, activation functions etc. in a specific neural network and mostly depend upon the users intuition. Radial basis function type neural network architecture is found to have the best approximation ability [29] to interpolate any finite data set in the n-dimensional parameter space. Thus for a given set of plant properties viz. gain ($K$), time constants ($\tau_1, \tau_2$) and delay ($L$), the RBF-NN is used to approximate the corresponding PID and FOPID parameters viz. $\{K_p, K_i, K_d\}$ & $\{K_p, K_i, K_d, \lambda, \mu\}$ respectively corresponding to a GA based optimal tuning which is a new concept and have not been investigated yet.

The RBFNN architecture in Fig. 1 consists of three layers, namely, the input layer, the hidden layer, and the output layer. The neurons in each layer are fully connected to the previous layer neurons. The input layer of the network is directly connected with the hidden layer of the network but connections between the hidden layer and the output layer are weighted linearly. The inputs are assigned to the neurons in the input layer directly and the outputs are also taken from the output layer neurons directly. Thus in general, the number of neurons in input and output layer is equal to the number of inputs and outputs respectively.

The nonlinear activation functions are placed in the hidden layer. The activation function used are the radial basis functions can be represented as $\phi(X(k)) \; \forall \; k \in \{1,2,...,N\}$ ($N$ being the number of inputs), where $\|.\|$ denotes the Euclidean 2-norm and $\phi$ is a nonlinear function commonly taken as Gaussian RBF:

$$\phi(X(k)) = e^{\left(\frac{-\|X(k)-x(i)\|^2}{2\sigma_i^2}\right)} \; \forall \; i \in \{1,2,...,N\} \quad (9)$$

Here, $x(i)$ is the center of the i$^{th}$ basis function in the i$^{th}$ hidden neuron, dimension being the same as the input vector $X(k)$ and $\sigma_i$ is called its radius or spread. The value of the spread parameter $\sigma_i$ is taken as 1. The j$^{th}$ RBF network output can be then represented as:

$$y_j(k) = w_{0,i} + \sum_{j=}^{M} w_{i,j} \varphi(X(k)) \quad (10)$$

where, $w_{ij} \; \forall \; i \in \{1,2,...,N\}; \; j \in \{1,2,...,M\}$ are the weights connecting the hidden neurons to the output neurons and $w_{0,i}$ is the weight connecting the bias to the output neurons.

Neural networks based on RBF is then used to map the relationships between reduced order process parameters $\{K, L, \tau_1, \tau_2\}$ as inputs and controller parameters $\{K_p, K_i, K_d\}$ and $\{K_p, K_i, K_d, \lambda, \mu\}$ as outputs for PID and FOPID respectively by minimizing the Mean Square Error (MSE) between the optimal GA tuned controller dataset and the RBFNN output.

Fig. 2 shows the schematic for the online gain and order scheduling of the Fractional Order $PI^\lambda D^\mu$ controller with a time varying process. The supervisor senses the process and extracts the corresponding reduced order SOPTD model parameters i.e. $\{K, L, \tau_1, \tau_2\}$. These are fed to the trained neural network, which adjusts the gain and order of the $PI^\lambda D^\mu$ controller. The sensor and actuator dynamics of the control loop are neglected in the present simulation.

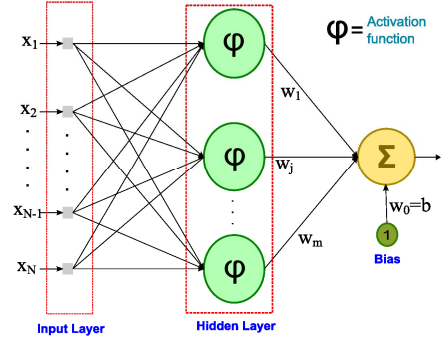

Fig. 1. Schematic representation of the RBF Neural Network.

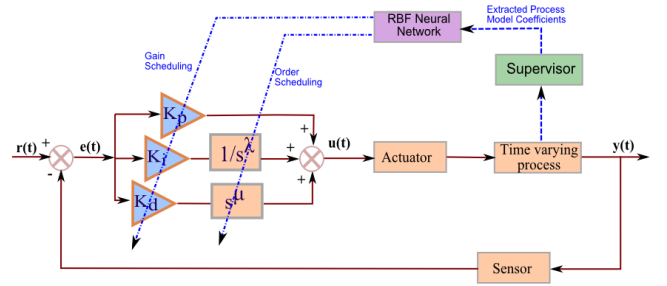

Fig. 2. Schematic of gain and order scheduled FOPID with RBFNN.

## IV. RESULTS AND DISCUSSIONS

### A. RBF NN Trained Controller Parameters

The GA based sub-optimal reduced order SOPTD parameters and the optimal PID and $PI^\lambda D^\mu$ controller parameters are now used to train the RBF neural network that can handle change in process model variation, represented as the switching within the test-bench of processes (3)-(6). The following Figs. 3-10 indicate using blue-dots, the discrete controller values for each family of processes $P_1, P_2, P_3, P_4$ which are obtained by the GA based tuning method minimizing the objective function (8). The RBFNN is used to fit this non-linear mapping between the plant parameters and the controller values producing optimum set-point tracking and minimum control effort requirement. It is evident from Fig. 3-10 that the red lines pass through all the blue dots, indicating that the RBF fitting is very accurate at the discrete tuned values and can also interpolate for plant parameters, lying in between the maximum and minimum bounds of the process parameter.

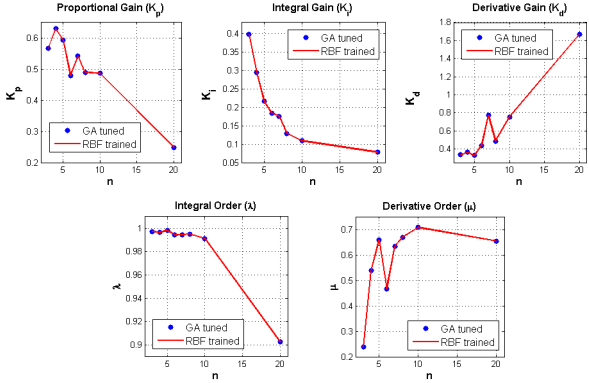

Fig. 3. Optimal FOPID controller parameters for process $P_1$.

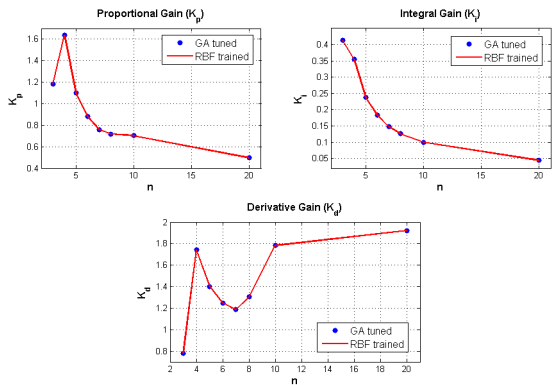

Fig. 4. Optimal PID controller parameters for process $P_1$.

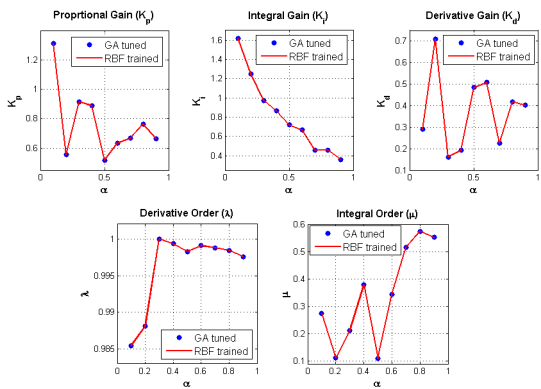

Fig. 5. Optimal FOPID controller parameters for process $P_2$.

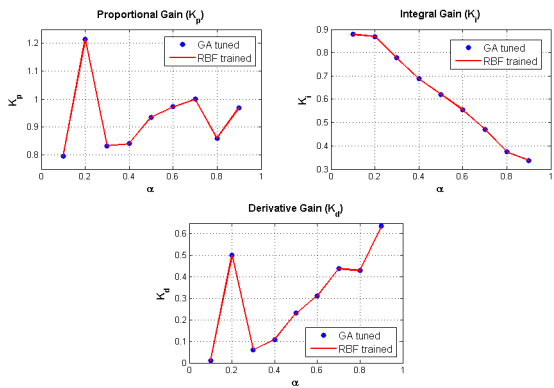

Fig. 6. Optimal FOPID controller parameters for process $P_2$.

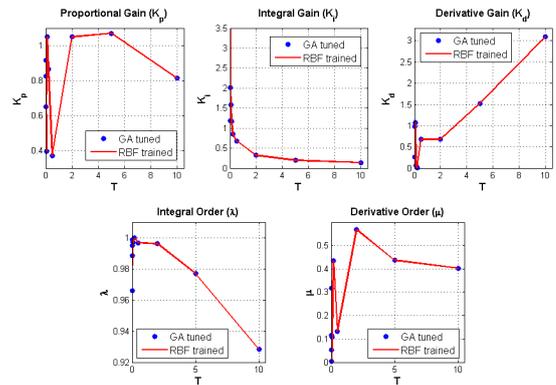

Fig. 7. Optimal FOPID controller parameters for process $P_3$.

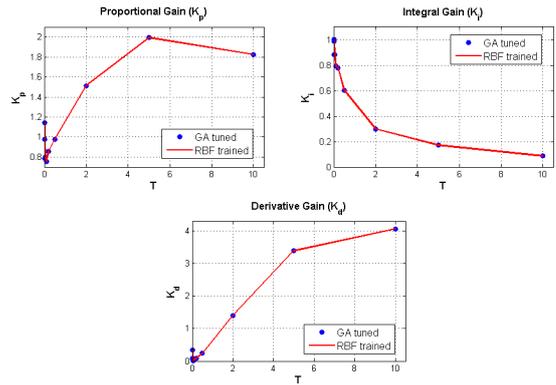

Fig. 8. Optimal PID controller parameters for process $P_3$.

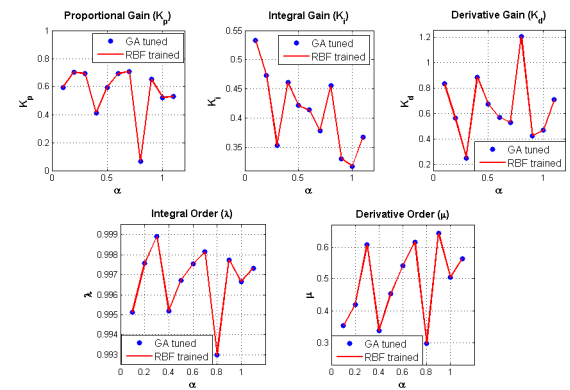

Fig. 9. Optimal FOPID controller parameters for process $P_4$.

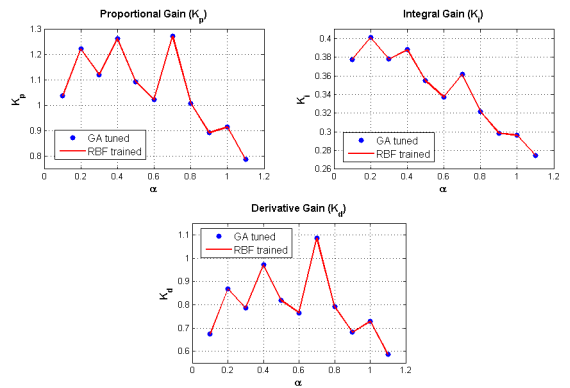

Fig. 10. Optimal PID controller parameters for process $P_4$.

## B. Gain and Order Scheduling of Optimal $PI^\lambda D^\mu$ Controllers for Random Switching Between the Test-Bench of Higher Order Processes

The online gain scheduling for different process models is simulated with the help of MATLAB and Simulink blocks. The limit within which the set point can vary, the frequency of the change of set point as well as the distribution which the magnitude of set point will follow can be pre-specified by the user. In the developed model, the process gets changed randomly according to a given distribution. The ANN block receives the process parameters at every simulation step and dynamically produces the optimum PID/FOPID controller gain and integro-differential orders.

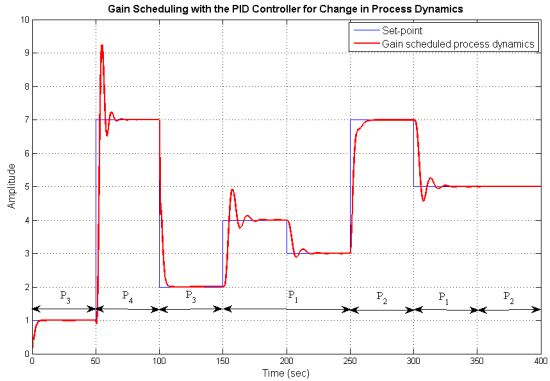

Fig. 11. Gain scheduling of the PID controller for switched process dynamics.

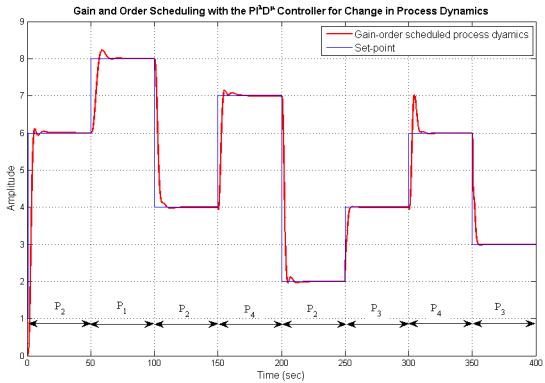

Fig. 12. Gain and order scheduling of the $PI^\lambda D^\mu$ for switched processes.

Figs. 11 and 12 show the gain scheduling of PID and gain-order scheduling of FOPID controllers respectively for arbitrary set-point tracking of an arbitrarily time varying process. The loop alternates randomly between the processes $\{P_1, P_2, P_3, P_4\}$ which can be thought of a linear time varying (LTV) process for the purpose of demonstration. The magnitude of set-point change and the time of occurrence are also changed randomly as soon as the process changes, for better visualization of the proposed gain and order scheduled $PI^\lambda D^\mu$ controller action. Stabilization and control of switched LTI processes have been discussed in [30]-[33], in a detailed manner.

As is evident from Fig. 11 and 12 even though the PID and $PI^\lambda D^\mu$ controllers are optimally tuned for each process for a unit change in set-point, they exhibit sharp transients when there is a shift in reduced order process model and consequent update in the controller parameters. Hence for applications which are sensitive to such jerks, a hierarchical supervisor may be used for monitoring system performance as mentioned before. However it is also evident that the transients in the PID controller are much higher than the $PI^\lambda D^\mu$ controller when the controllers are scheduled to get updated with corresponding process model and hence the FOPID controllers are more adept at suppressing these transients and consequently better suited for scheduling over conventional PID controllers [1], even with intelligent supervision [12], [34]-[42].

## V. CONCLUSIONS

Simulation for online scheduling of controllers for time varying processes has been done by RBF Neural Networks. A wide variety of test bench processes have been taken into consideration to represent the dynamics of an arbitrary time varying process which can not be optimally controlled with fixed controller parameters. The higher order process models are converted to reduced-order SOPTD template and are tuned by Genetic Algorithms minimizing a weighted error index (ITAE) and weighted squared control signal. The non-linear relationship between the SOPTD process parameters and PID/FOPID controller parameters that ensures optimum set-point tracking with minimum control effort are mapped by the RBF Neural Network and is shown to give an excellent fit. The trained RBF neural network is used for online scheduling of the PID parameters (gain) and FOPID controller parameters (gain and order) for an arbitrarily time varying process. Simulation results show that the FOPID controllers are better at suppressing switching transients and are consequently better suited over their integer order counterparts for such gain scheduling based adaptive control methods. Future scope of research can be directed towards other available FO adaptive control techniques [43]-[47] and online scheduling FO controller design for more accurate and complicated reduced parameter models [48].